\documentclass[twocolumn,showpacs,preprintnumbers,amsmath,amssymb]{revtex4}

\usepackage{graphicx}
\usepackage{dcolumn}
\usepackage{bm}

\begin{document}


\title{Coevolution of game and network structure: The temptation increases the cooperator density}

\author{Shao-Meng Qin}
\affiliation{Institute of Theoretical Physics, Lanzhou University, Lanzhou $730000$, China}

\author{Guo-Yong Zhang}
\affiliation{Institute of Theoretical Physics, Lanzhou University, Lanzhou $730000$, China}

\author{Yong Chen}
\altaffiliation{Corresponding author. Email: ychen@lzu.edu.cn}
\affiliation{Institute of Theoretical Physics, Lanzhou University, Lanzhou $730000$, China}

\date{\today}

\begin{abstract}
\textit{Most papers about the evolutionary game on graph assume the statistic network structure. However, social interaction could change the relationship of people. And the changing social structure will affect the people's strategy too. We build a coevolutionary model of prisoner's dilemma game and network structure to study the dynamic interaction in the real world. Based on the asynchronous update rule and Monte Carlo simulation, we find that, when players prefer to rewire their links to the richer, the cooperation density will increase. The reason of it has been analyzed.}
\end{abstract}


\maketitle

\section{\label{introduction}introduction}

Cooperation is a key aspect in the real world, ranging from biological systems to human behavior ~\cite{nature1,nature2}. Therefore, people restore to the game theory to study the emergency and maintenance of cooperation in biology, psychology, computer science, and economics~\cite{biology,book1,book2,PR}. Especially, the prisoner's dilemma game (PDG), has become a metaphor to approach the emergency of cooperation and altruism behavior. In the tradition PDG, each of two players chooses a strategy from cooperation ($C$) or defection ($D$) simultaneously and gets payoff. They both receive $R$ upon mutual $C$ and $P$ upon mutual $D$. A defector gets $T$ when it plays game with cooperator who gets $S$. In PDG, we have $T > R > P > S$ and $2R > S+T$. Because the mutual $C$ get the highest total income, $D$ is the better choice than $C$ no mater what the other player's strategy. Without any mechanism for the evolution of cooperation, natural selection favors defection. The other widely studied games include snowdrift game~\cite{SNG,SNG2}, public good game~\cite{PGG}, rock-paper-scissors game~\cite{RPS}, and so on.

The complex network has also attracted lots of attentions in the past few years. The complex network is ubiquitous in nature. The human society can also be described as the systems composed of interacting agents. The classical social network maps the individual into the node, and the connection between individuals into the link. The evolutionary game theory in spatial structure has became a unifying paradigm to study how cooperation may be sustained in a structured population~\cite{Nowak}. It was found that the spatial extension is one of several natural mechanisms to enforce cooperation. Network structure will affect the behavior of strategy density~\cite{structure}. In lattice network, the cooperation is usually get together to support each other to resist the defection~\cite{lattice1,lattice2,SNG2}. Santos and Pacheco found in Scale-Free networks the strong correlation leads to the dominating trait throughout the entire range of parameters of both games in scale-free networks~\cite{SF}. And also, there are anmount of researches on other networks, like small-world~\cite{SW} and random network~\cite{random}.

When the player on the structure network chose the better strategy to play game, in fact, not that the players select the proper strategy, but player's strategy is determined by the network structure. For example, in scale-free networks, the large degree nodes (hubs) and the nodes which connect to hubs tend to be occupied by $C$~\cite{SF}.

The networks used in the most papers of this field are statistic. The connection will never change once it is build. It is not realistic enough, as the interactions themselves help shape the network~\cite{SW}. What is more, in the real world, the relationship between the people is not constant. Sometimes people cannot cut some relationship with their relatives, neighbors or colleagues but they can end their old relationship and build a new one. Sometimes this changing is caused by the results of the game, because people would like to make friends in a reciprocal respect. For example, people always like to make friends with rich one for a sake of pursuing fortune. So, when we study the social model in network like PDG, the network structure should be dynamical entities~\cite{Arne}. The nodes can remove or sustain their link in network according to the game results.

Till now, there are few models studied the cooperative behaviors in a groups with adaptive connections. Besides some early work~\cite{eW1,eW}, Arne build a coevolution model of strategy and structure~\cite{Arne}. In this model, the probability of forming or cutting link between node $A$ and $B$ is based on their strategies. The changing of network structure is result from the strategy changing in the network. Then it also affect the strategy density back. However, the link could change even if the nodes' strategy do not change in their model. The rewire of link in this model is not the player's own decision. Li \textit{et al.} also build a coevolution model that the node rewire its link only for changing its strategy~\cite{LRL}. Moreover, in this model, the node rewire its long range link based on the existed network structure, not the playing game results.

In our opinion, a rational model for coevolution of game and network structure should contain two features: (1) The nodes rewire their links only when agents change their status; (2) The rewiring should be based on the playing results of game. In this paper, we will present a coevolution model of the PDG and network. We use PDG as a metaphor to studying cooperation between unrelated individuals and consider a social networks with four fixed local links and one adjustable long-range link (LRL). The agents in the network play game with their network neighbors. They will change their strategies and adjust LRLs according to the results of game. Then the network structure changing also affect the cooperation density.


\section{\label{model}Model}

We set up a system of $N$ players arranged at the nodes of a ring lattice network. Each node is connected with four local nodes. These local interactions will not change during the whole process of the evolution. Besides four fixed links, every node in this lattice has an adjustable LRL which connects to another node and self-connections and the duplicate links are excluded. We call the LRL out-link for the node to whom it belong or in-link for the node to whom it connect. The node can select another node to which the out-link wires, but it cannot give up the LRL. Therefore, each node has at least one out-link and many possible in-links. When node changes its strategy, it will also rewire its LRL. We will discuss when and how LRLs rewire later.

As suggested by Nowak and May~\cite{Nowak}, we adopt $R=1$, $T=b$ $(1<b<2)$, and $S=P=0$. Then $b$ can be considered as the temptation to $D$ against $C$. Every player plays the PDGs with its neighbors on network and itself and get the total payoff $W$. After each round of the game, players are allowed to inspect their neighbors' total payoffs and change their strategies in the next round. The player $i$ updates its strategy by selecting one of its neighbors $j$ with a probability $\gamma _{ij}$,
\begin{equation}
\gamma_{ij} = \sum_{m \in \Omega_i} \frac{k_j (t)}{k_m (t)},
\label{eq1}
\end{equation}
where $\Omega_i$ is the community composed of the nearest neighbors of the player $i$, and $k_m(t)$ is the degree of node $m$ at time $t$. In the spirit of preferential attachment proposed by A.-L. Bar\'abasi and R. Albert~\cite{PS}, we incorporate the preferential selection rule to model social behaviors. In Eq.~\ref{eq1}, player with large degree has more probability to impact his neighbors. That is true in the society that people who have great impact often have lots of social relations and they are also focused by their friends. Node $i$ will follow the node $j$'s strategy by the probability,
\begin{equation}
W = \frac{1} {1 + \exp \left[ (W_i-W_j)/\kappa \right]},
\label{eq2}
\end{equation}
where $W_i$ and $W_j$ are the total payoffs of node $i$ and $j$, and $\kappa$ indicates the noise generated by the players allowing irrational choices ~\cite{ka1,lattice1,lattice3}.

If node $j$ has the same strategy with $i$ or $i$ do not mimic $j$'s strategy, node $i$ will do nothing. Otherwise, it will rewire its LRL to a new one. There are two rewiring rules in our model: random rewiring and preferential rewiring. With probability $p_c$, the density of cooperation in the network, node $i$ will chose a new node randomly. For the rest probability $1-p_c$, node $i$ will chose a new node according to the node's payoff. In the preferential rewiring rule, the node rewires its link according to the payoff of all nodes in network,
\begin{equation}
\lambda_{ij}=\sum_{m \in G} \frac{W_j^{\alpha}}{W_m^{\alpha}},
\label{eq3}
\end{equation}
where $\lambda_{ij}$ is the probability of node $i$ rewiring its link to $j$ and $G$ presents all nodes in the graph. $\alpha$ is used to change the effect of payoff. $\alpha=0$ indicates that the payoff has no effect here and the nodes rewire their links randomly. For $\alpha>0$, the node will prefer to connect the node with larger payoffs. So it also looks like a kind of preferential selection rule.

\section{\label{results}Simulation Results}

We run our simulations with varying $b$ and $\alpha$ for fixed $\kappa=0.1$ and the system size $N=1000$. All the results in this paper are obtained from the average results with $100$ different Monte Carlo (MC) simulation trails. We start with node linking its LRLs to other nodes randomly with equal probability and random initial state with $p_c=0.5$ as the initial state. The players update their strategies in random sequence. In every MC step, all nodes have one chance to change their strategies and rewire their links.

\subsection{\label{A}Strategy evolution }

\begin{figure}
\begin{center}
\includegraphics[width=0.5\textwidth]{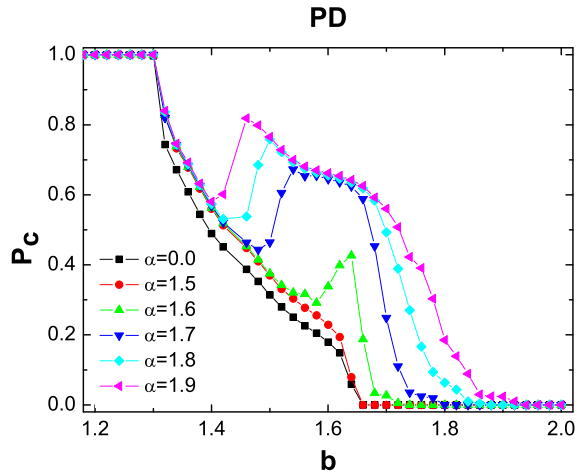}
\caption{\label{fig1}(Color online) Frequency of cooperators $p_c$ for different $\alpha$ as functions of the advantage of defectors $b$.}
\end{center}
\end{figure}

\begin{figure}
\begin{center}
\includegraphics[width=0.5\textwidth]{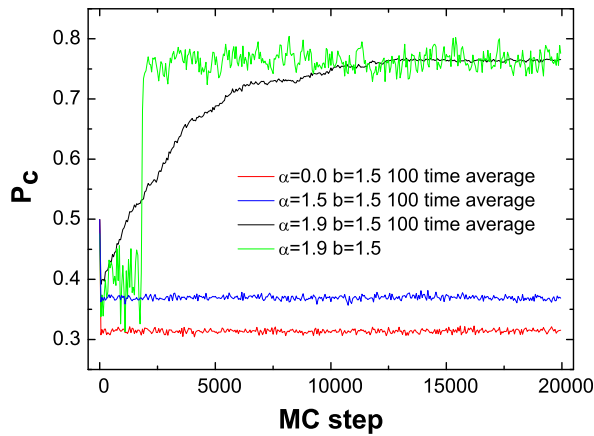}
\caption{\label{fig2}(Color online) Frequency of cooperators $p_c$ evolve with $t$ for systems at different parameters on PDGs.}
\end{center}
\end{figure}

Figure~\ref{fig1} shows the frequency of cooperators $p_c$ in our model as the functions of $b$ for different $\alpha$. Similar to evolutionary game in regular network~\cite{lattice1,lattice2}, we also find two thresholds in our model. Full cooperation is achieved if $b$ does not exceed the threshold $b_{c1}$. For $b>b_{c2}$, $C$ cannot resist the temptation of $b$ and cannot survive in the network. In the region of $b_{c1}<b<b_{c2}$, $C$ and $D$ can coexist in the network. Compared with the case of $\alpha=0$, the position of $b_{c1}$ does not change with $\alpha$. However, $\alpha$ affect the $b_{c2}$ conspicuously. 

The probability of node using preferential selection to rewire its LRL is $1-p_c$. Therefore $\alpha$ does not work at $p_c$ close to $1$ or $b$ close to $b_{c1}$. When $\alpha<1.6$, the qualitative results $p_c$ remain unaffected by $\alpha$ that $p_c$ decreases monotonous with $b$. When $\alpha>1.6$, there exists a region of $b$ promoting cooperation obviously. This promotion starts at $b=1.64$ ($\alpha=1.6$) and this region enlarge with increasing $\alpha$. But the effect of promotion does not increase with $\alpha$. We observe that $p_c$ does not change at $1.55<b<1.65$ for $\alpha=1.7$, $1.8$, and $1.9$. Actually, the transition is caused by the changing of network structure. We will discuss it in the next subsection.

In order to discuss how the $\alpha$ promotes $p_c$ in the promotion region, we present the time evolutions of $p_c$ in Fig.~\ref{fig2} for fixed $b=1.5$ with different $\alpha$ values. The red, blue, and black lines are the averages of $100$ trials for $\alpha=0$, $1.5$, and $1.9$ respectively. The green one is the $p_c$ time series of one trail in the black line. For $\alpha=0$ and $1.5$, $p_c$ decreases with time to its station state quickly. As shown in Fig.~\ref{fig1}, $p_c$ for $\alpha=1.5$ is a little higher than that of $\alpha=0$. However, for $\alpha=1.9$, $p_c$ decreases like $\alpha=0$ firstly, and then the evolution of network drives $p_c$ increasing with time to $0.76$. Considering that the black line is the average of $100$ trails, we believe the green line in Fig.~\ref{fig2} contains more details of the evolution. In the early stage of the green line, $p_c$ decreases to a temporary stable state in a manner similar to but a little larger than $\alpha=1.5$. However, at $t=2000$, there is a sharp increasing in the green line from about $0.4$ to $0.76$ which is also the final level of the average result (the black line). It means that the gradually increasing of the black line is caused by the average effect of $100$ same sharply increase at different times.

\subsection{\label{B}Network structure}

In this model the behavior of $p_c$ and the evolution of network structure are equal important. The evolution of network structure results in the transition of $p_c$.

\begin{figure}
\begin{center}
\includegraphics[width=0.5\textwidth]{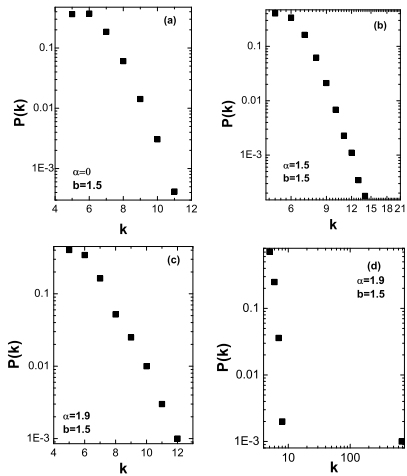}
\caption{\label{fig3}(Color online) The cumulative stationary degree distributions $P(k)$ in PDGs.}
\end{center}
\end{figure}

In order to describe the network structure, we first present the degree distribution $P(k)$ in Fig.~\ref{fig3}. Panel (a) is $P(k)$ in the case of the stable state of red line in Fig.~\ref{fig2}. Here the preferential rewiring does not work and all LRLs select the target nodes randomly. Considering the self-connection is forbidden, we know
\begin{equation}
P(k) = C_{N-5}^{k-5} \left( \dfrac{1}{N-4} \right) ^{k-5} \left( 1-\dfrac{1}{N-4} \right) ^{N-k+5}.
\nonumber
\end{equation}
Here $N$ usually is large enough, so one can get
\begin{equation}
P(k) = C_{N}^{k-5} \left( \dfrac{1}{N} \right) ^{k-5} \left( 1-\dfrac{1}{N} \right) ^{N-k+5}.
\nonumber
\end{equation}
Figure~\ref{fig3}(b) is $P(k)$ for the stable state of blue line in Fig.~\ref{fig2}. $P(k)$ in (b) is similar to that of (a) but the largest degree is $19$. Fig.~\ref{fig3}(c) is $P(k)$ for the stable state of gree line in Fig.~\ref{fig2} and (d) is for the green line after the sharp increasing. 

\begin{figure}
\begin{center}
\includegraphics[width=0.3\textwidth]{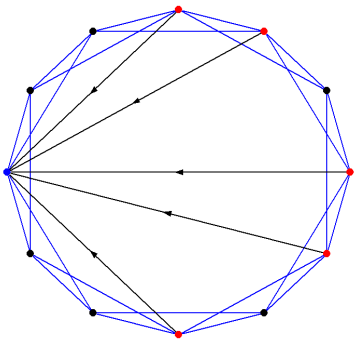}
\caption{\label{figbc}(Color online) Illustration of HN, AN, and BN. Each node in the network has four fixed links and there are five red nodes wire their LRLs to the blue one. In order to make AN and HN prominent, we do not draw the LRLs of other nodes. The blue node has the largest degree in this net, so blue node is HN, and the red one is AN and the others are BN. We draw the arrows in the figure to present these LRLs are out-links for AN and in-links for HN.}
\end{center}
\end{figure}

Both (c) and (d) in Fig.~\ref{fig3} are the degree distributions of one trial, but not the cumulative stationary degree distribution of $100$ different trials. By comparing (c) with (d), it is helpful to uncover the reason of the sharp increasing in Fig.~\ref{fig2}. In Fig.~\ref{fig3} (d), there is only one node that its degree is larger than half of the other nodes connected to it. We name this node which has the largest degree in the network as hub node (HN). As presented in Fig.~\ref{figbc}, the other nodes can be divided into two types: the nodes connect their LRLs to HN and the nodes do not. We name the first node as AN and the second one as BN. The number of them are $N_A$ and $N_B$, respectively.

Now, we exam the detail of the network after the sharp increasing in the green line ($\alpha=1.9$, $b=1.5$) of Fig.~\ref{fig2}. Note that the strategy of HN is always $C$ and the strategy of most ANs is also $C$. Before the sharp increasing or in the case of other parameters without sharp increasing, the HNs  are also prefer to $C$. This phenomenon is also observed in some other networks with hub nodes~\cite{SF,LRL}. More detailed information of our model are listed in Table~\ref{tab1}. 

In Table~\ref{tab1}, $p_{Ac}$ is the cooperation density of AN and $p_{Bc}$ is for BN. Almost all nodes of ANs chose the strategy $C$, so we do not need to present the mean payoff of AN with $D$. What is more, it is found that $p_{Bc}=0.308$ is close to the case of $\alpha=0$ ($p_c=0.314$ for $b=1.5$, $p_c=0.235$ for $b=1.55$, and $p_c=0.179$ for $b=1.6$). It means that the existence of AN does not affect the strategy density of BN. As discussed in Ref.~\cite{LRL}, AN can resist the temptation of $b$ by mimicking the strategy of HN. After the sharp increasing, the probability that AN mimics the strategy of HN is much larger than that of other neighbors. The HN's payoff is also larger because it has a lots of in-link LRLs. We will discuss the details of these probabilities in the next subsection. On the other hand, only the node with strategy $C$ can grow into HN. If HN is occupied by $D$, HN will get higher payoff temporarily. However, as we discussed above, AN will follow HN's strategy and the strategy of AN will be $D$. Then the HN cannot earn payoff from its in-link LRLs. Once HN cannot earn enough payoff, both preferential and random rewiring will drive ANs to rewire its LRL to other nodes. Then a new HN with strategy $C$ will appear in the network. So it seems that strategy $C$ is a better choice for HN because it can earn a stable higher payoff. 

From Table~\ref{tab1}, we also find that the BNs with $D$ earn the most payoff and the payoff of BN with $C$ is close to the payoff of AN. However, although the mean payoff of BN with $D$ is the highest, in fact, the density of cooperator doesn't decrease with time. It shows that the probability of $C$ mimicking $D$ strategy and $D$ mimicking $C$ strategy are the same.

\begin{table}
\caption{The detailed information of prisoner's dilemma games ($\alpha=1.9$).}
\begin{ruledtabular}
\begin{tabular}{c|c c c}
            & $b=1.5$   & $b=1.55$  & $b=1.6$\\ \hline
  $N_A$       & $669$     & $621$     & $605$\\
  $N_B$       & $330$     & $378$     & $394$\\ \hline
  $p_c$    & $0.766$   & $0.686$   & $0.663$\\
  $p_{Ac}$     & $0.992$   & $0.988$   & $0.987$\\
  $p_{Bc}$     & $0.308$   & $0.191$   & $0.164$\\ \hline
  payoff of AN& $5.213$   & $4.806$   & $4.682$\\
  payoff of BN with $C$& $5.222$ & $4.987$   & $5.025$\\
  payoff of BN with $D$& $5.713$ & $5.541$   & $5.561$\\
\end{tabular}
\end{ruledtabular}
\label{tab1}
\end{table}

\begin{figure*}
\begin{center}
\includegraphics[width=1.2\textwidth]{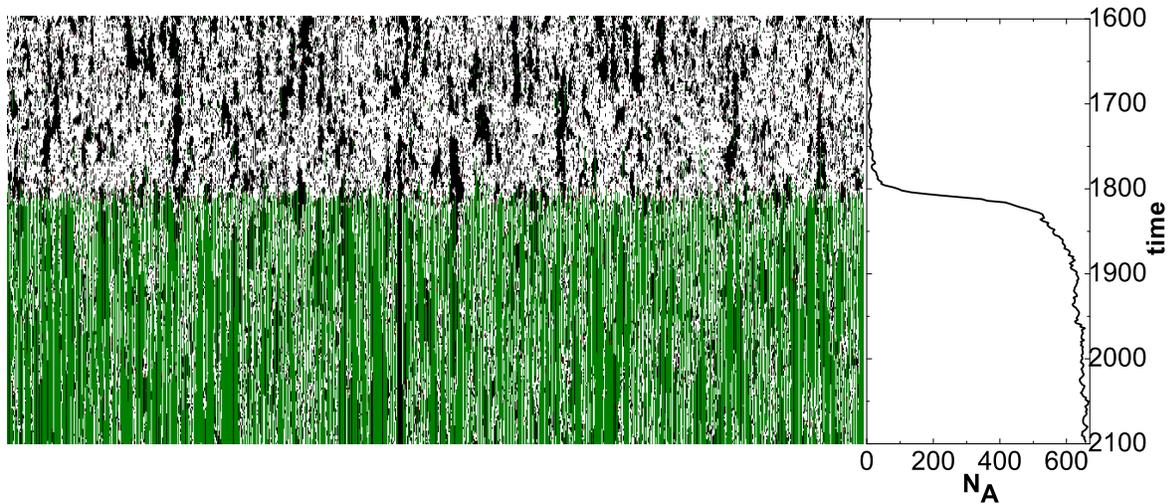}
\caption{\label{fig4}(Color online) Because the network structure in our model is one dimension lattice, we can use a color line to present the snapshot of the status of the network. The black and white dots present $C$ and $D$ in BN. The green and red dots present $C$ and $D$ in AN. In order to know how the AN, BN, and strategy evolve in network, we arrange these snapshots with time from top to the botton at $1600<t<2100$ in the left panel for the green line in Fig.~\ref{fig2}. The right panel presents the time evolution of the number of AN at the same time.}
\end{center}
\end{figure*}

Each horizontal line in Fig.~\ref{fig4} presents a snapshot of the network. We arrange these snapshots with time from top to the botton to show how the AN evolves with time. So we can depict every player’s strategies in network and observe the evolution of these strategies. The riht panel is $N_A$ at the same time with the left. There is also a sharp increasing of $N_A$ at the same time like the green line in Fig.~\ref{fig2} in looks. At $t=1750$, about $50$ MC steps before the transition, $N_A$ increase gradually from about $10$ to $50$. After the sharp increasing, $N_A$ still increase gradually to the final stable state. Moreover, before the sharp increasing happened, one can observe many black blocks (the upper part of the left panel in figure~\ref{fig4}). It means the model has the similar feature of PDG in regular network that the $C$ node tends to get together for blocks to resist the $D$. These blocks start at a few $C$s, maybe three or more, and then close to each other in the network coincidentally. Then a block is established and it will grow to change their neighbors' strategies. After some MC steps, the block will shrink and then disappears in the last. After the sharp increasing of $N_A$, there are too few red dots ($D$ in AN). The green strip ($C$ in AN) indicates that the ANs or BNs are very stable in the network. The probability of AN change to BN is very small and vice versa.

\subsection{\label{Analyse}Discussion}

Based on the results in the above context, the effect of $\alpha$ is different from various $b$ and $\alpha$. After the sharp increasing, the nodes in network can be divided into AN and BN. Almost all AN are $C$ and the density of $C$ in BN is close to the case of $\alpha=0$. So we can use the mean field theory and some basic feature of stable state to explain why the sharp increasing happened.

After the sharp increasing in Fig.~\ref{fig3}, the system reaches the stable state gradually. Then we have ${d N_A}/{d t}=0$ or $N_{A\rightarrow B}=N_{B\rightarrow A}$, where $N_{A\rightarrow B}$ is the average number of nodes changed from AN to BN in one MC step and $N_{B\rightarrow A}$ is that changed from BN to AN.

Considering that there are too few $D$s in AN, we assume that $N_{A\rightarrow B}$ is only caused by $C \rightarrow D$ and random rewire. Here, we neglect the preferential rewiring. Because the contribution of preferential rewiring is only about $2\%$ of random rewiring. Then we get
\begin{eqnarray}
N_{A\rightarrow B} &=&(1-p_c) Q_{A \rightarrow B} p_c N_A \nonumber \\
&&\frac{ \left( 5+\frac{N-N_A}{N} \right) \left( 4+\frac{N-N_A}{N} \right)}{N_A+ \left( 5+\frac{N-N_A}{N} \right) \left( 4+\frac{N-N_A}{N} \right)}.
\label{eq4}
\end{eqnarray}
Here, $p_c$ means the change happened in the random rewiring, and $5+(N-N_A)/N$ is the mean degree of nodes in networks. We neglect self-connection and multi-connection forbidden and we have $N \approx N-1$ here. Because AN has the same strategy with HN, AN only mimics the strategy from other $4+(N-N_A)/N$ neighbors. The big fraction is the probability of AN do not chose HN to mimic the strategy. The last $(1-p_c)$ is the probability of mimicked target with strategy $D$. We assume $Q_{A \rightarrow B}$ is the probability of success in the mimicking.

Then $N_{B\rightarrow A}$ will be more complicate. We assume that BN change to AN because they use the preferential rewiring. The contribution of random rewiring is about $0.2\%$ of preferential rewiring, so we neglect it and derive the following formula, 
\begin{widetext}
\begin{eqnarray}
N_{B\rightarrow A} &=& \left[ p_{Bc}(1-p_c)+(1-p_{Bc})p_c \right] Q_{B\rightarrow A} (1-p_c)N_B  \nonumber \\
&&\frac{(N_A+5)^{\alpha}}{(N_A+5)^{\alpha}+N_A (2+4p_c)^{\alpha}+ p_{Bc} N_B \left( 1+(5+\frac{N-N_A}{N})p_c \right) ^{\alpha} + N_B (1-p_{Bc}) \left( b(5+\frac{N-N_A}{N})p_c \right) ^{\alpha}},
\label{eq5}
\end{eqnarray}
\end{widetext}
where $1-p_c$ means the preferential rewiring, and $p_{Bc}(1-p_c)$ is the probability of BN with strategy $C$ to mimic its $D$ neighbor and that $D$ try to mimic its $C$ neighbor. The fraction here is the probability of node rewire to HN using the preferential rewiring. We assume $Q_{B \rightarrow A}$ is the probability of success in the mimicking.

Now, one can get $p_{Bc}$ from the simulation of $\alpha=0.0$ and $p_c=(p_{Bc}N_B+N_A)/N$, and
then we know how $N_A$ evolves with time by using $N_{B\rightarrow A}- N_{A \rightarrow B}$. If $N_{B\rightarrow A}- N_{A \rightarrow B} = 0$, $N_A$ will not change with time. And $N_{B\rightarrow A}- N_{A \rightarrow B} > 0$ means $N_A$ will increase in the next MC step. However, we do not know $Q_{A \rightarrow B}$ and $Q_{B \rightarrow A}$ yet. According that the mean payoff of $BN$ with $D$ is larger than mean payoff of $AN$ in Tab.~\ref{tab1}, we conjecture $M = Q_{B \rightarrow A} / Q_{A \rightarrow B}$ and $M>1$. Indeed, we find $M=2.0$ is fit to our model. We will take $b=1.5$ with $\alpha=1.9$ and $\alpha=1.3$ as examples. For $b=1.5$ and $\alpha=0$ we get $p_c=0.314$ from the simulation. 

\begin{figure}
\begin{center}
\includegraphics[width=0.5\textwidth]{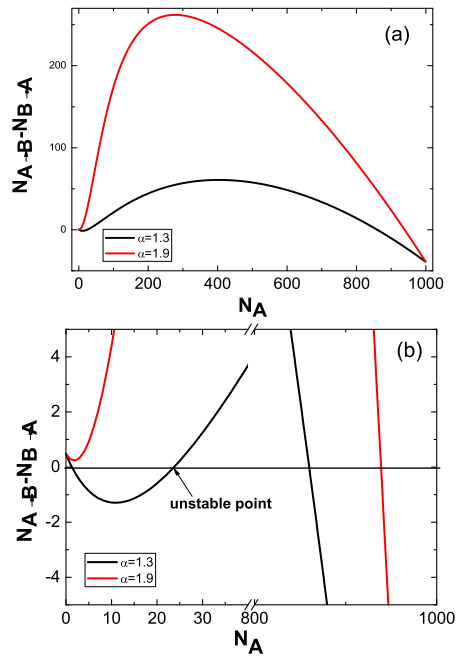}
\caption{\label{fig5}(Color online) $N_{B\rightarrow A}- N_{A \rightarrow B}$ with various $N_A$. Panel (b) is an enlargement of (a).}
\end{center}
\end{figure}

Fig.~\ref{fig5} plots ${d N_A}/{d t}$ as different $N_A$. For $\alpha=1.3$ there are two stable points at $N_A = 3$ and $860$, and one unstable point at $N_A = 23$. For $\alpha=1.9$, there is only one stable state at $N_A=939$. The unstable point will decrease with the increasing of $\alpha$ and coincident with the first stable point at $\alpha=1.69$. However, even $\alpha=0$, the maximal degree in the network is about $12$. So the first stable point can be discarded. When the unstable point crosses $N_A=12$ or there is only one stable point, the system will reach to the second stable point.

\section{\label{results}Conclusion}

The coevolution of dynamics and network structure is rapidly becoming an important field of the evolutionary game. It contains more details about the social interaction in the real world. In this paper, we build a co-evolution model of PDG and network structure. Each node in network has four fixed local links and one adjustable LRL. When the node changes its strategy, it will rewire its LRL to another node according to the node's payoff and density of cooperation. And we introduce a parameter $\alpha$ to denote the effect of payoff.

Many early works~\cite{eW1,ew,Arne,LRL} also proved that the adaptive network can enhance the cooperation. All these enhancements are caused by the emergency of cooperator with large degree in the network. In~\cite{eW1}, the cooperation is very sensitive to the plasticity parameter and only the adaptive network can enhance the cooperation. 

In our model, the players rewire their LRLs for any $\alpha$, but the cooperation is enhanced only in the case of $\alpha>0$ that this enhancement is obvious for $\alpha>1.6$ in a certain region of $b$. However, our results show that the enhancement of cooperation only happen in the case of changing the network structure property. In our model, for $\alpha=0$, the node will also rewire its LRL, but the network property will not change and the cooperation level will not be enhanced. The cooperation is enhanced only when the node rewires its LRL according to the payoff. Similar phenomena was also observed in our simulations with snowdrift game (SG). We found that SG is more sensitive to $\alpha$ than PDG and the obvious enhancement is for a smaller $\alpha$. So we conjecture that the coevolution of network structure and game is an important mechanics to maintain the cooperation in the real society.

Different from the results in~\cite{eW1,eW,Arne,LRL} which the cooperation always dominates in the adaptive network and the increasing of cooperation is limited. That is caused by two reasons: (1) In the probability of $p_c$ the player use the random rewiring. (2) The existance of four fixed links in network can be regarded as a noise to prevent the preferential selection. In~\cite{eW}, authors discussed the leaders and the global cascades. If every node could change its strategy in a smaller probability, the global cascades of coopertation is also observed in our model.

The analysis in this paper is based on the balance of AN and BN. However, when the sharp increasing didn't happen, perhaps there exist more than one HNs and HN is changing from one node to another frequently. Because of the absense of the information about spacial structure and Eq.~\ref{eq2}, the presented analysis in this model is not very precise any more. Actually, it is impossible to include all the details for the analysis. We just hold on the main factors of the model and it works well enough to explain the main features of our model.

\begin{acknowledgments}
This work was supported by the National Natural Science Foundation of China under Grant No. $10305005$ and by the Fundamental Research Fund for Physics and Mathematic of Lanzhou University. This study is supported by the high-performance computer program in Lanzhou University.
\end{acknowledgments}

\end{document}